# Fold-switching proteins

Devlina Chakravarty[1] and Lauren L. Porter[1,2,*]

[1]National Center for Biotechnology Information, National Library of Medicine, National Institutes of Health, Bethesda, MD 20894
[2]Biochemistry and Biophysics Center, National Heart, Lung, and Blood Institute, National Institutes of Health, Bethesda, MD, 20892

Email addresses: devlina.chakravarty@nih.gov, porterll@nih.gov

Running title: Fold-switching proteins

*Correspondence:
National Library of Medicine
8600 Rockville Pike
Building 38A, Room 6S608
Bethesda, MD 20894
301-827-0924
porterll@nih.gov

**Abstract**

Globular proteins are expected to assume folds with fixed secondary structures, α-helices and β-sheets. Fold-switching proteins challenge this expectation by remodeling their secondary and/or tertiary structures in response to cellular stimuli. Though these shapeshifting proteins were once thought to be haphazard evolutionary byproducts with little intrinsic biological relevance, recent work has shown that evolution has selected for their dual-folding behavior, which plays critical roles in biological processes across all kingdoms of life. The widening scope of fold switching draws attention to the ways it challenges conventional wisdom, raising fundamental unanswered questions about protein structure, biophysics, and evolution. Here we discuss the progress being made to answer these questions and suggest future directions for the field.

# 1. Introduction

For over 50 years, globular proteins been expected to assume single folded structures fostering their biological functions (7). Here we describe an emerging class of proteins that defy this expectation. **Fold-switching proteins**, once thought to be transitory evolutionary intermediates with little intrinsic biological relevance (105, 116), have been increasingly observed to regulate biological processes (55) by remodeling their secondary and/or tertiary structures in response to cellular stimuli (83). Further, both folds of many fold-switching proteins have been selected by evolution, indicating that their shapeshifting behavior confers advantage (95). To date, nearly 100 fold switchers have been characterized experimentally (83). Examples span all kingdoms of life, perform many biological functions, and respond to numerous triggers. By examining common features of fold-switching proteins, it has been estimated that up to 4% of proteins in the Protein Data Bank (PDB) switch folds (83). A subsequent survey estimates that up to 5% of *E. coli* proteins may switch folds (59).

Though fold-switching proteins represent a relatively small fraction of the protein universe when compared to canonical single-folding proteins or intrinsically disordered proteins (IDPs), they challenge conventional wisdom and raise fundamental unanswered questions about protein structure, biophysics and evolution. For example:

- Conventional wisdom says that a protein's primary sequence encodes its unique structure (7); fold-switching proteins' primary sequences encode at least two stable structures (95). How does one sequence encode two conformations with distinct secondary and/or tertiary structures? How do proteins interconvert between these structures?
- Conventional wisdom says that homologous proteins assume similar structures (90, 100); fold switching demonstrates that homologous proteins can assume different structures (19, 117). What distinguishes homologous sequences that encode similar structures from those that encode different ones?
- The decades of conventional wisdom that laid the foundation for the Nobel Prize-winning AlphaFold model (50) are not enough to predict fold switching reliably (16, 17). What makes fold-switching proteins so difficult to predict?

Perhaps by answering these questions, we will gain a more fundamental understanding of protein structure, biophysics, and evolution that can be applied more broadly. This review presents the progress being made to answer the questions that fold-switching proteins pose.

## 2. One sequence, two folds

Many proteins change their conformations by undergoing rigid body motions (24) or local rearrangements (52). **Fold-switching** proteins change their conformations by remodeling their secondary and/or tertiary structures in response to cellular stimuli. Indeed, the ~100 fold-switching proteins currently known were largely identified by searching the Protein Data Bank (PDB) for protein structures with identical (or near-identical) sequences, regions with different secondary structures, and a clear functional justification for the structural differences (83). The secondary and tertiary structures of fold-switching proteins can be remodeled in numerous ways, some larger and some smaller (15). Three of the best characterized fold switchers are shown in **Figure 1**.

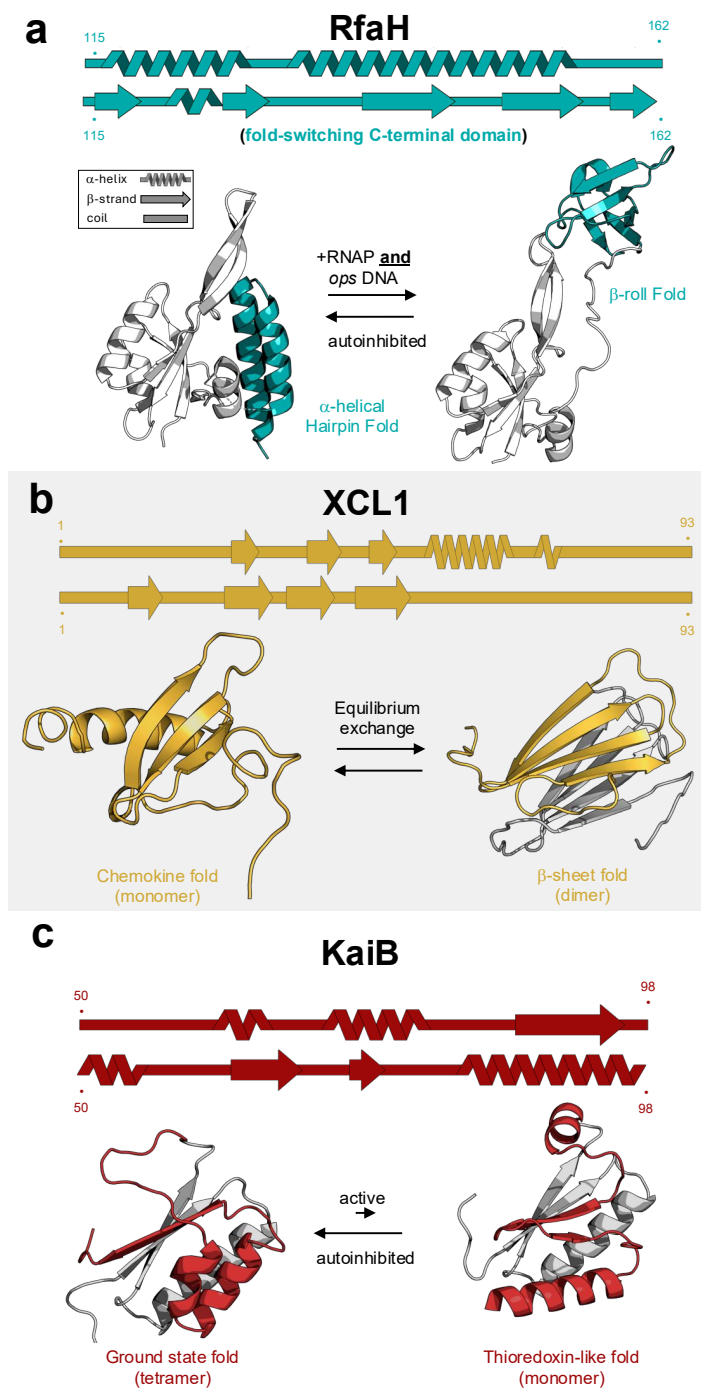

**Figure 1.** *Three well-characterized fold-switching proteins.* (**a**) The C-terminal domain (CTD, teal) of *E. coli* RfaH reversibly switches from an α-helical to β-sheet fold upon binding RNA polymerase and operon polarity suppressor DNA. Adapted from (82). (**b**) Under physiological conditions, the human chemokine XCL1 (mustard) reversibly interconverts between a chemokine fold involved in signaling and a dimeric β-sheet fold that binds bacterial and fungal pathogens. (**c**) The C-terminal subdomain of *S. elongatus* KaiB (red) switches secondary structures when its ground state tetrameric form dissociates into a monomer and binds the circadian clock protein KaiC. Single-folding regions of all proteins are colored gray. Secondary structures diagrams were generated using SSDraw (21) with corresponding residue numbers above and below; all ribbon diagrams in this paper were generated with PyMOL (97).

- RfaH is a member of the universally conserved NusG family of transcription factors. Bacterial NusGs, including RfaH, have two domains: an N-terminal domain (NTD) that binds RNA polymerase, enabling transcription elongation, and a C-terminal domain (CTD) that typically folds into a Kyprides, Ouzounis, Woese (KOW) β-roll fold. Unlike most of its NusG homologs with solved structures, RfaH's C-terminal domain folds into an α-helical hairpin fold in its unbound state, masking the RNA-polymerase binding site of its NTD (**Figure 1a**). Upon binding RNA polymerase and a specific DNA sequence called operon polarity suppressor (*ops*), RfaH's CTD dissociates from its NTD (123), unmasking its RNA-polymerase binding site. The dissociated CTD reversibly refolds into the KOW β-roll (124), which binds an integral subunit of the ribosome, fostering both efficient transcription and translation elongation.

- XCL1 is a human chemokine involved in both signaling and pathogen response. Under physiological conditions it reversibly interconverts between a monomeric chemokine fold that activates the G-protein coupled receptor XCR1, mediating influx of intracellular calcium ions, and a dimeric fold that binds glycosaminoglycans (107) and pathogenic bacteria (77) and fungi (29) (**Figure 1b**). Though its secondary structure changes are less drastic than RfaH's and KaiBs, XCL1's interconversion involves completely reregistering its hydrogen bonded network of β-sheets and repacking its hydrophic core (30). Each conformation is sampled approximately equally under physiological conditions (107).

- KaiB is an essential component of the cyanobacterial circadian oscillator, which can be reconstituted *in vitro* with just KaiB, two other proteins, ATP, and $Mg^+$ (76). KaiB largely assumes a ground state fold, whose C-terminal half assumes a βααβ fold, but it exchanges with a monomeric thioredoxin-like fold whose C-terminal half assumes an αββα fold (106). The cyanobacterial clock is deactivated by knocking out KaiB's fold switching (20), highlighting its importance in this biological process. KaiB's slow interconversion rate is important to the clock's 24-hour cycle (121). Proline isomerization contributes to this slow rate; mutating the prolines that isomerize to other amino acids speeds the interconversion (114, 121).

The conformational changes presented in **Figure 1** represent the dramatic remodeling of secondary and/or tertiary structure that fold-switching proteins can undergo. Other fold switchers can undergo more localized fold-switching events, such as a single α-helix to β-hairpin transition that archaeal selecase undergoes upon oligomerization (68). Biophysical factors that enable fold switching are discussed in the next section.

## 3. Biophysical properties of fold-switching proteins

*3.1. Protein energy landscapes*

Protein structure spans a stability spectrum ranging from highly stable folded proteins to unstable **intrinsically disordered proteins (IDPs)**. **Energy landscapes** illustrate the fundamental linkage between protein stability and conformational homogeneity. The energy landscapes of many stable single-folding proteins contain a single deep energy well corresponding to their unique thermodynamically stable conformations. While non-functional, off-pathway intermediates can exist, they are expected to represent minor, difficult-to-detect species in the protein's native ensemble (57). By contrast, IDPs lack a well-defined energy minimum; instead, they populate broad, shallow basins or exist as dynamic, extended ensembles (18).

Fold-switching proteins differ from both single folders and IDPs (32, 61). Their energy landscapes feature multiple minima (**Figure 2**), each corresponding to a structurally defined, native-like conformation. These alternative folds are folded and biologically relevant. Some are formed by irreversible transitions stabilized by a change in environment, such as membrane insertion of hemolytic pore proteins (9, 103), while the alternative folds of **metamorphic proteins** (75) are metastable and capable of reversible interconversion under physiological conditions. While this behavior challenges the classical "one sequence–one structure" paradigm (31), some suggest that fold switching remains compatible with Anfinsen's principle, provided all populated states represent local or global energy minima accessible under biologically relevant conditions (109, 110).

The multi-conformational nature of fold switchers comes with an energetic cost: they often exhibit marginal thermodynamic stability, with folding free energies ($\Delta G_{fold}$) sometimes greater than −3 kcal/mol, significantly less stable than the −15 to −5 kcal/mol range observed for most globular proteins (18, 101, 121, 122). This low stability facilitates access to alternative structures that comprise a small but meaningful fraction of the population at equilibrium (>1% for $\Delta G_{fold} \geq -2.6$). These alternative structures can be disparate from the dominant structure under a given set of conditions, such as α-helix ⇌ β-sheet conversions or shifts in β-sheet register, which reorganize a large fraction of the protein's tertiary contacts within a domain or folding unit.

Recent work suggests that cold denaturation may play an important role in regulating the relative populations of fold-switched conformations (67). Specifically, one conformation may be stable at low temperature but unstable at high, while the other may be stable at higher temperatures but unstable at low. Thus, changing temperature can shift the conformational equilibrium from one dominant conformation to another (101, 107, 121). This hypothesis may apply broadly to single-domain fold switchers that interconvert at equilibrium; its applicability to proteins that switch in response to binding partner or domain cleavage remains to be seen.

*3.2. Experimentally characterized mechanisms of interconversion*
How do fold-switching proteins transition between the relatively shallow wells of their energy landscapes? Recent work has supported a previous proposal: fold-switching protein regions can often unfold and refold independently from the larger body of the protein (83), accessing partially unfolded intermediates that bridge distinctly folded structures. Since conventional structure determination techniques—particularly X-ray crystallography—are inherently biased toward the most thermodynamically stable or highly populated conformation and provide a static rather than dynamic picture of fold switching, interconversions of fold switchers are typically observed using nuclear magnetic resonance (NMR) spectroscopy techniques.

Two of the NMR techniques most used to observe conformational exchange of fold switchers are chemical exchange saturation transfer (CEST, (108)) and zz-exchange (35). The former examines conformational changes on the order of milliseconds to seconds, the latter on the order of seconds. Sub-millisecond-to-millisecond motions can be observed with Carr-Purcell-Meinbloom-Gill (CPMG) relaxation dispersion experiments (78). All three techniques were recently combined to probe conformational exchange in the C-terminal domain (CTD) of RfaH, which reversibly interconverts between all α-helix and all β-sheet folds (**Figure 1a** (14)). Though the all-helix conformation has only been observed in the presence of RfaH's N-terminal domain, its isolated CTD–which predominantly assumes its β-sheet structure–was found to populate four minor states, three of which involved smaller conformational changes that exchanged on the order of milliseconds. The fourth most populated minor state assumed an α/β hybrid structure that exchanged with its major all-β structure on the order of seconds. This intermediate structure contains elements of both of RfaH's dominant forms, suggesting that it may play an important role in the CTD's α-helix ⇌ β-sheet transition.

A clear trend has emerged from applying NMR techniques to several other fold-switching proteins: their large conformational changes occur on the order of seconds or slower. This has been observed through zz-exchange experiments on the 3-α-helix bundle ⇌ α/β-plait transition of the temperature-sensitive fold switcher Sa1 V90T (101) and the α/β to all-β transition of XCL1 (30). Temperature-dependent NMR experiments have shown that KaiB, a fold switcher that regulates the timing of the cyanobacterial circadian clock (20), switches on the order of hours and populates a partially disordered state (114, 121). Further $^{19}$F-NMR experiments show that the transition between the active and inactive conformations of the *Mycobacterium tuberculosis* protein PimA also occurs on the order of seconds (62).

*3.3. Molecular dynamics simulations*

Though classical molecular dynamics (MD) simulations and **Markov models** have elucidated biologically important protein dynamics (12), they struggle to accurately simulate fold switching because of the relatively slow timescale on which it occurs (seconds or slower). MD simulations can be computationally expensive and may not adequately sample large-scale structural rearrangements that occur on the order of seconds or longer, while Markov models might oversimplify the complex nature of fold-switching (111). Alternative techniques are being developed to address these challenges and provide a more realistic understanding of fold-switching processes, such as **replica exchange MD (REMD)**(40), **replica exchange with tunneling (RET)**(10), and **simplified structure-based models (SBMs)**(85).

*3.3.1. Structure-Based Models of Protein Folding Landscapes*

Structure-based models (SBMs) offer a simplified yet powerful approach for exploring protein energy landscapes. In the single-basin formulation, the complex conformational space is reduced to a dominant energy funnel that guides the system toward a single native state. This model type is especially useful for studying protein folding thermodynamics and kinetics, as it captures essential folding features while maintaining computational tractability. The free energy surface, defined over atomic or residue-level coordinates, often contains many local minima corresponding to metastable conformations, which can complicate convergence toward the global minimum in simulations. To address this, Seifi and Wallin introduced an energy term based on the native contact map, effectively deepening the energy funnel toward the folded structure and enhancing the model's ability to reproduce the biologically relevant conformation (98). This refinement improves simulation convergence and enables more accurate characterization of folding pathways and transition state ensembles (85).

While single-basin SBMs are well-suited to proteins with a single stable structure, fold-switching proteins—those capable of adopting multiple native states—require more elaborate modeling. These cases require a multi-basin SBM, which includes separate energy terms to bias the system toward each of the alternative native conformations (**Figure 2**). Such models enable investigation into the determinants of conformational plasticity and the energy transitions underlying fold switching events. Refolding pathways of several biologically significant fold-switching proteins have been successfully characterized using both all-atom and coarse-grained implementations of single- and dual-basin structure-based models (SBMs). Notable examples include the human chemokine XCL1, which switches between a chemokine-like fold and a dimeric β-sheet structure(53); the spindle checkpoint protein Mad2, which toggles between open and closed conformations to regulate cell cycle progression (41); the cytolytic toxin ClyA (42), which undergoes a transition from a soluble monomer to a membrane-inserted pore-forming oligomer; and viral proteins such as influenza hemagglutinin (63) and the SARS-CoV-2 spike protein, both of which display large conformational shifts essential for membrane fusion and host cell entry (42). Dual-basin SBMs have also captured the conformational transitions of the transcriptional regulator RfaH, whose C-terminal domain (CTD) transitions from all-α to all-β folds (38, 85, 99, 123). Further, for the cyanobacterial circadian clock protein KaiB, they indicated that the dissociation of dimers plays a crucial role in its structural change (20). These insights directed experiments: size-exclusion chromatography and hydrogen-deuterium exchange mass spectrometry (HDX-MS) were conducted on the R75C KaiB mutant, which accelerates the cyanobacterial clock components by approximately 2 hours. The analysis showed that this mutant exists in both dimeric and monomeric forms, and HDX-MS results showed increased local flexibility in areas corresponding to the fold-switched (fsKaiB) conformation. These experimental results supported the computational model, establishing a clear connection between dimer dissociation and fold switching in KaiB (86).

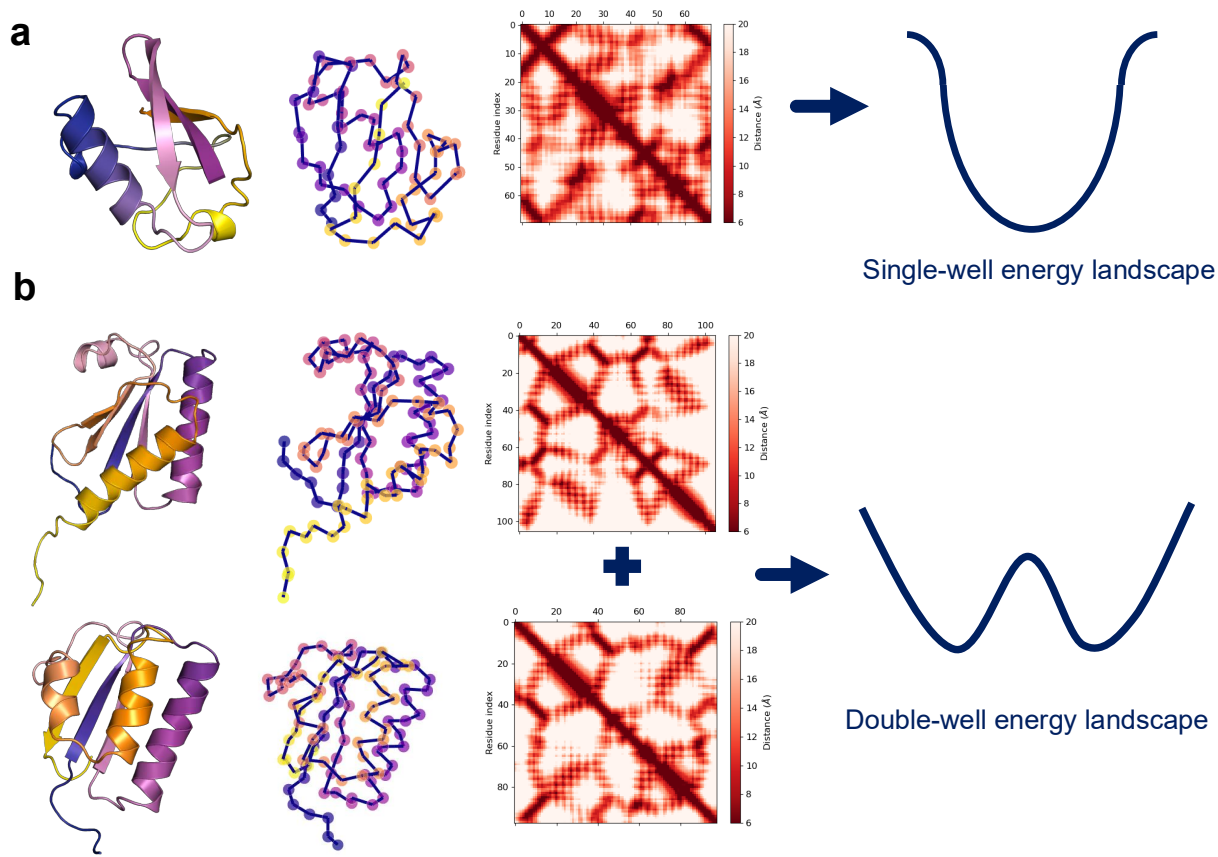

**Figure 2.** *Schematic representation of energy landscapes modeled by structure-based models (SBMs).* (a) In single-basin SBM, the energy landscape is shaped as a funnel biased toward a single native conformation, typically derived from the contact map of the folded structure; Ubiquitin is used as an example here (PDB ID: 3EHV). This enables efficient sampling of folding pathways and thermodynamic properties. (b) For metamorphic proteins such as KaiB (PDB IDs: 2QKE and 5JYT), which adopt multiple native states, the SBM must be generalized to a double-well model. This is achieved by incorporating separate energy terms, each favoring a different native topology, thereby allowing exploration of fold-switching dynamics and conformational equilibria. Energy terms are derived from reference structures. These structures can be obtained through experiments or simulation modeling.

*3.3.2 Extended Molecular dynamics (MD), Replica exchange with and without tunneling (RET), capture fold-switching landscape*

Replica-exchange simulated tempering, also known as parallel tempering, is a Monte Carlo simulation technique used to efficiently sample the configuration space of complex systems, such as proteins. In this approach, multiple replicas of the system are simulated in parallel at different temperatures, enabling exploration across a range of energy landscapes. Periodic exchanges of configurations between replicas allow lower-temperature simulations to overcome energy barriers by temporarily accessing higher-temperature states. Standard replica exchange improves sampling by facilitating transitions between local minima that would otherwise be rarely visited at low temperatures (70). Replica exchange with tunneling (RET) is an advanced variant designed to further accelerate sampling, particularly for systems with rugged energy landscapes like protein folding or aggregation. This method incorporates transitions not only between temperatures but also between different levels of system resolution, typically switching between coarse-grained and fine-grained representations. By enabling tunneling between resolutions, the system can traverse conformational barriers more efficiently, enhancing the exploration of rare but functionally relevant states (69).

Both methods have been successfully applied to investigate the folding landscapes of certain fold-switching proteins. Bernhardt et al. demonstrated the effectiveness of replica exchange with tunneling (RET) using a designed 11-residue peptide and two 56-residue variants representing the A (all-α) and B (α/β) domains of protein G (11). RET's enhanced sampling capabilities were further leveraged to map the free energy landscape of RfaH-CTD and propose a mechanism for its conformational conversion (68). Additionally, extended molecular dynamics (MD) simulations combined with principal component analysis of atomic fluctuations and thermodynamic modeling, based on both configurational volume and free energy landscape, have been used to characterize the conformational thermodynamics of human XCL1 and one of its reconstructed ancestors (120). These computational studies (53) provide insight into the proteins' thermodynamic landscapes, highlighting the importance of configurational entropy and the free energy surface within the essential space (i.e., the space defined by generalized internal coordinates showing the largest, often non-Gaussian, structural fluctuations).

## 4. Protein fold switching and evolution

### 4.1. Proteins with similar sequences but different folds

Decades of empirical observation indicates that proteins with similar amino acid sequences assume similar folds (26, 90). To test the limits of this observation, Creamer and Rose introduced the "Paracelsus challenge" (88, 89), questioning whether a protein's structure and function could switch if ≤50% of its sequence was changed. Dalal et al. met this challenge by designing a sequence that folded into a four-helix bundle resembling the Rop protein homodimer. This sequence was 50% identical to the predominantly β-sheet Bl domain of streptococcal protein G (25). Remarkably, the engineered protein retained native-like characteristics and was named Janus, after the Roman god with two faces, symbolizing its dual identity. Further, Jones et al. (49) partially transformed a small, disulfide-linked β-sheet protein into an α-helical hairpin, while Yuan and Clarke (119) achieved some success in converting an all-helical protein into one resembling the B1 domain of protein G, using a similar strategy to that of Dalal and colleagues.

Some of the most impressive work inspired by the Paracelsus challenge involves engineering two monomeric protein domains with up to 98% sequence identity and different folds and functions (2, 3, 46). Demonstrating that this switch could be triggered throughout the sequence, several single mutations that switched the protein between one fold (3-α-helix bundle) and the other (α/β-grasp) were later identified (45). More recently, this work has culminated in the design of a network of three proteins with 100% sequence identity but different folds (91), supported by previous work, which showed that embedding a small protein domain within a larger fold can force the small domain to assume an alternative fold (81).

Together, these findings challenge conventional wisdom by showing that proteins with highly similar sequences can adopt different folds and functions, presenting far-reaching implications. First, they point to a more fluid model of protein fold space in which highly similar–and even identical–sequences can assume very different folds and functions (79). Thus, proteins with similar sequences do not necessarily assume similar folds. Indeed, state-of-the-art artificial intelligence models, such as AlphaFold2, have confidently mispredicted structures of proteins that assume folds different from their homologs with solved structures (15). Second, these findings suggest that other biological processes can cause proteins to switch their folds and functions. Two cases have been observed:

i. *Alternative splicing.* The human oncoprotein BCCIP has two isoforms (α and β) that differ by one exon substitution. Though the sequences of these isoforms are 80% identical, they assume completely different folds that differ by >10 Å root-mean-square deviation (66) and have mutually exclusive binding partners. A conformational switch in the Piccolo C2A domain is also regulated by alternative splicing of a nine-residue sequence, changing not only its secondary structural elements but also its calcium binding affinity (39).

ii. ***Single-nucleotide polymorphisms***. In various forms of non-Hodgkin lymphoma, the most frequent mutation (D83V) of human protein myocyte enhancer factor 2B (MEF2B) induces an α-helix to β-sheet fold switch. This structural switch is believed to alter the DNA-binding function of MEF2B and/or interactions with proteins involved in multiple signaling pathways, possibly explaining MEF2B's involvement in oncogenesis (60).

Additionally, ***post-translational modifications (PTMs)*** and ***changes in codon translation rates*** may also induce fold switching. While experimental evidence for these events triggering a fold switch is lacking, it remains probable (79). Indeed, phosphorylation has been suggested as a possible trigger for the fold switching of Orf9b, a SARS-CoV-2 protein (43).

*4.2. Roles of fold-switching proteins in evolution*

Fold-switching proteins can be either transitory evolutionary intermediates as one fold evolves into another (116, 117) or evolutionary end products selected for their specific functions (30, 95). Their roles can be traced through evolutionary trajectories characterized by phylogenetic analysis and ancestral sequence reconstruction (32, 33). Dual-fold selection can be identified by evolutionary couplings uniquely corresponding to the different conformations of fold-switching proteins (95). Notably, if fold switching is an adaptive trait (i.e., the result of selective pressure), fold-switching proteins might not be as scarce as they appear in databases of solved structures (32). This section describes how fold switchers have been observed both as evolutionary intermediates and as selected evolutionary end products (**Figure 3**).

Though new protein folds have generally been proposed to evolve from random sequences (92) or by mixing and matching fixed elements of secondary structure (4, 56), recent work has shown that new folds can emerge from secondary structure transitions (α-helices ⇔ β-sheets) engendered by stepwise mutation, a process called **evolved fold switching** (19). This process was first suggested in the Cro transcription factor family (58, 87), but strong evolutionary inferences were impeded by limited sequence information and few solved structures. These technical barriers were recently overcome when winged helix (wH) folds were found to evolve from helix-turn-helix (HTH) folds in a large family of approximately 600,000 bacterial response regulators with 85 experimentally determined structures (19). Evolutionary relationships among family members were mapped through phylogenetic analysis, and putative evolutionary intermediates that may have bridged the two fold families were identified. **Ancestral sequence reconstruction (ASR)**, combined with AlphaFold2 structure predictions (50), suggested how stepwise mutations may have facilitated the transition from the ancestral helix-turn-helix to the winged helix fold, expanding DNA binding specificity (**Figure 3**). Another recent analysis tracked the evolutionary transition between two distinct β-barrel folds, one a core domain in RNA polymerases and the other in the ribosome (117). Through ancestral reconstruction, this study found a putative evolutionary intermediate that interconverted between both folds in response to small molecules such as phosphate and citrate.

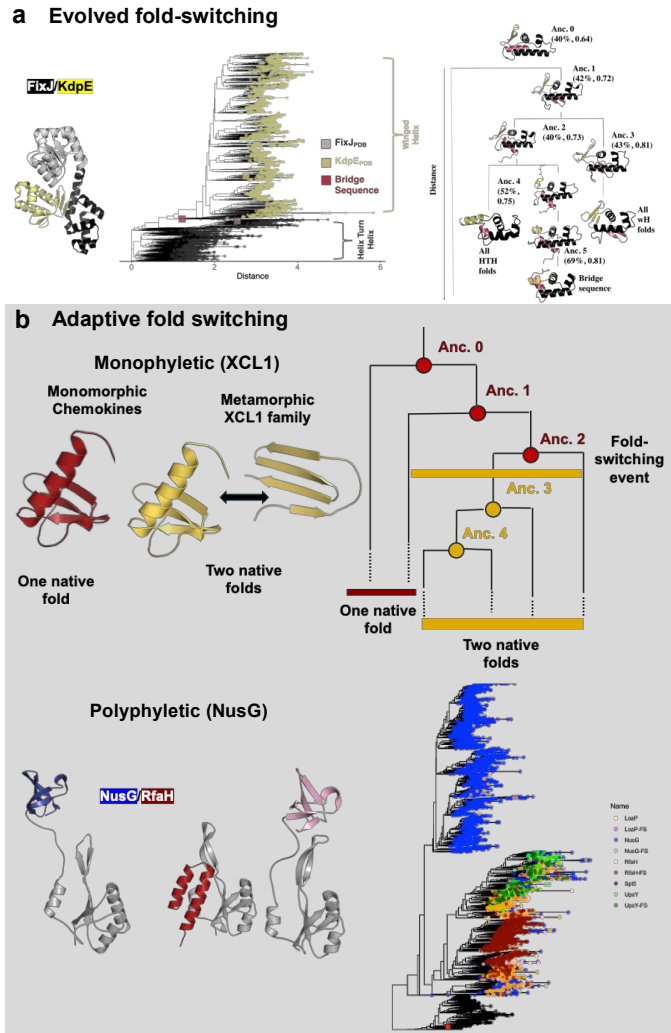

**Figure 3**. *Evolution of fold switching in protein families* (a) Evolved fold switching is illustrated by helix-turn-helix (HTH) response regulators evolving to winged-helix (wH) folds. FixJ (black) is an example of a HTH response regulator; KdpE (yellow) exemplifies a wH fold (left). Maximum-likelihood phylogenetic trees suggest an evolutionary path between response regulators with HTH and wH folds. Sequences with C-terminal domains annotated as HTH/wH from NCBI protein records are gray/yellow. The clade with 12 sequences bridging the two folds is highlighted in pink. AlphaFold models suggest that stepwise mutations through the bridge sequence caused HTH proteins to switch to wH. Distance units in both trees are arbitrary, though sequences further in space have more distant evolutionary relationships. (B) Adaptive fold switching can be either monophyletic (XCL1) or possibly polyphyletic (RfaH). **Monophyletic Fold-Switching:** fold switchers all came from a single common ancestor. This simplified phylogenetic tree traces XCL1's evolutionary path from its last common ancestor (Anc.0), which possessed the canonical chemokine fold, to two distinct subfamilies highlighted in red and yellow. The red subfamily retains only the chemokine fold, while the yellow subfamily can adopt both the chemokine fold and an alternative fold. Dishman et al. propose that fold-switching in XCL1 allows the protein to adapt its function to physiological needs: at sites of infection, the dimeric alternative fold helps fight bacteria, while the monomeric chemokine fold activates leukocytes via the XCR1 receptor, thus providing two functions without requiring the synthesis of a new protein or fold. The emergence of fold-switching (indicated by a yellow bar on the tree) was detected by analyzing ancestral sequences (Anc.0 to Anc.4) that were resurrected using ancestral sequence reconstruction (ASR). Access to a second native-state structure arose after the loss of a conserved disulfide bond (Anc. 2) and the accumulation of mutations that either destabilized the chemokine fold or favored the alternative fold (yellow). (C) **Polyphyletic Fold-Switching**: In contrast, polyphyletic fold-switching refers to multiple, independent occurrences of fold-switching in the RfaH/NusG family, which contains highly diverse sequences. In this tree, NusG branches that are non-switching folds are highlighted in blue, while the RfaH fold-switching branch is shown in red. The phylogeny was constructed using Maximum Likelihood analysis of ~6,000 unique sequences from the NusG/RfaH family, with branch support validated by bootstrapping. The large red square marks Q57818 in the clade containing all archaeal sequences (shown in black) and represents an Spt5 protein used as a reference for profile realignment in the multiple sequence alignment. The tree is rooted between the predominantly non-fold-switching NusG subfamily (blue) and the fold-switching clades.

Fold switching can also be an adaptive trait selected for the functional advantages it confers (95). This was first observed in the human chemokine XCL1 (30), a signaling protein that reversibly interconverts between a chemokine fold that binds the G-protein coupled receptor XCR1 and a dimeric β-sheet fold that binds fungal pathogens (**Figure 1b**). The metamorphic transcription regulator, RfaH, is another example where fold switching likely evolved from a single-folding precursor (**Figure 1a**). Unlike XCL1, whose fold switching seems to be monophyletic (evolved from a single common ancestor), fold-switching RfaH homologs appear to have evolved from multiple single folding ancestors (polyphyletic) and serve specialized cellular functions (**Figure 3**) (8). Fold switching may also be a universally conserved trait in some protein families, such as in the bacterial protein MinE (28) and the eukaryotic protein Mad2 (41, 47). Both families are involved in cell division, and evolutionary couplings unique to both folds are observed throughout both families (95).

Homologous protein sequences often–but not always–share a common function and fold. Consequently, large sets of homologous sequences (multiple sequence alignments, or MSAs) often present correlated mutational patterns due to evolutionary couplings (72) that usually correspond to amino acid pairs in direct contact (118). The exponential increase in sequenced genomes often allows for the ready detection of these evolutionary couplings and the inference of three-dimensional folds from predicted contacts. This leaves an open question: can a protein's proclivity towards conformational heterogeneity be observed from sequence variation in its MSA? In the study by Sutto et al., structural and dynamic properties of the catalytic domain of SRC tyrosine kinase were recovered by extracting evolutionary information from its MSA (104). Other works have hinted that amino acid contacts unique to each conformation of fold-switching proteins may have coevolved, a phenomenon also referred to as dual-fold coevolution (37, 82). Schafer et al. demonstrated that evolution has selected both conformations of many fold-switching proteins, suggesting that they likely confer a selective advantage (95). The study identified dual-fold coevolution in 56 known fold-switching proteins across various diverse families. It also provides a biological rationale for finding dual-fold contacts in fold-switching protein families, emphasizing the fact that **dual-fold coevolutionary signals** originate from the sequences of protein **subfamilies** populated by fold-switching proteins, rather than from **superfamilies** typically dominated by single-fold proteins.

These studies highlight the importance of identifying primary sequence features unique to fold-switching proteins, which differ from those of single-fold proteins. This understanding could facilitate a systematic search for metamorphic candidates within the proteome. Furthermore, pinpointing proteins on the cusp, where small mutations might trigger fold switching, could reveal sequence traits that improve structural flexibility. Such insights would support the rational design of fold-switching proteins and deepen our understanding of how certain disease-related mutations impact protein function (80).

## 5. New fold switchers through prediction and design

*5.1. Predicting fold switchers from their sequences*

Current bioinformatics techniques for predicting protein structure often fail to recognize fold-switching proteins. Further, molecular dynamics-based methods are too computationally expensive to model individual fold switching transitions without prior knowledge of both conformations, much less search for fold switchers systematically on a large scale. Consequently, a significant number of fold-switching proteins likely remain uncharacterized. To address this gap, computational tools that reliably predict fold switching behavior directly from sequence data are needed. Developing these tools would help to elucidate the sequence–structure relationship, define the biological scope of fold switching, and show how variations in sequences can lead to different conformational states.

Despite their many impressive capabilities, state-of-the-art AI-based structure prediction models, such as AlphaFold2 (AF2), AlphaFold3 (AF3 (1)), AF2-based enhanced sampling techniques (93), diffusion-based models (48), and **large language models** such as ESMfold (64) fail to consistently capture both conformations of fold-switching proteins (16, 17, 27, 48). Though these models can predict some protein structures by coevolutionary inference, they fail to accurately predict fold switched conformations even when their coevolutionary signals are clearly present in the input MSA (95). For example, AF2 consistently predicts a helical structure of RfaH despite strong evolutionary couplings corresponding to its β-sheet configuration (17). Removing RfaH's helical structure from AF2's training set and retraining causes the model to predict the expected β-sheet fold (96). This indicates that AF2 sometimes memorizes structures in its training set and associates them with MSA sequences rather than inferring structures from coevolutionary signals (94). Structure memorization explains why approaches like AF-cluster (113) work in limited cases, such as certain KaiB variants, but systematically fail on fold-switching proteins (17). Since AF-cluster fails completely when using a version of the AF2 architecture without fold-switched conformations in its training set (13, 96) and random sequence sampling greatly outperforms AF-cluster when running the original AF2 model (59, 94), we conclude that AF-cluster works by associating memorized structures or substructures with MSA sequences, not coevolutionary inference of input MSAs as originally claimed (113). Recent work suggests that these associations occur through conservation patterns (59), explaining AF2's limited ability to generalize fold-switching predictions among diverse protein homologs (15, 17, 82). A similar issue occurs with AF3, which incorrectly assigns coevolutionary restraints to the dimeric protein XCL1, while AF2 correctly predicted its structure, likely due to its learned representations and misassignment of coevolutionary signals (17).

Nevertheless, several approaches indicate that fold-switching proteins can sometimes be identified through conservation patterns. Several methods have successfully inferred fold switching from inconsistent or uncertain secondary structure predictions based on statistical profiles of MSAs (22, 54, 73, 74). One of these methods–based on the secondary structure predictor JPred4 (34)– successfully predicted fold switching in many sequence-diverse homologs from the universally conserved NusG transcription factor family (82). The method predicts that 24% of sequences in this family undergo large $\alpha$-helix $\rightleftharpoons$ $\beta$-sheet transitions. Predictions were consistent with circular dichroism and nuclear magnetic resonance spectroscopy experiments for 10/10 sequence-diverse variants, many of which had pairwise sequence identities <35%. Extending predictions to all NusG homologs suggested that that fold switching may be a pervasive mechanism of transcriptional regulation in all kingdoms of life.

Though AlphaFold-based methods do not reliably predict fold switching, random sequence sampling leveraged by a method called CF-random has been shown to improve prediction success and efficiency for certain fold switchers, likely by leveraging conservation patterns (94). CF-random outperforms all other AlphaFold-based methods for predicting multiple conformations of not only fold switchers and fold-switched assemblies but also other proteins that undergo rigid body motions and local conformational rearrangements (59). Notably, it achieves higher success with 3- to 8-fold fewer samples than competing methods. This increased efficiency is important since CF-random generates 200 predictions/target (less than the 1000+ sampled by other methods (51, 113)). Large pools of targets require subsequent filtering to reach the correct solution(s). This is especially important when the experimental structures are not available to validate the predictions empirically. Thus, a correct prediction when scored poorly may be lost in the sample of unphysical models. It should be noted that, while CF-random currently outperforms other predictive approaches for known fold-switchers, it achieves only a 35% success rate, underscoring the continued challenges in this field.

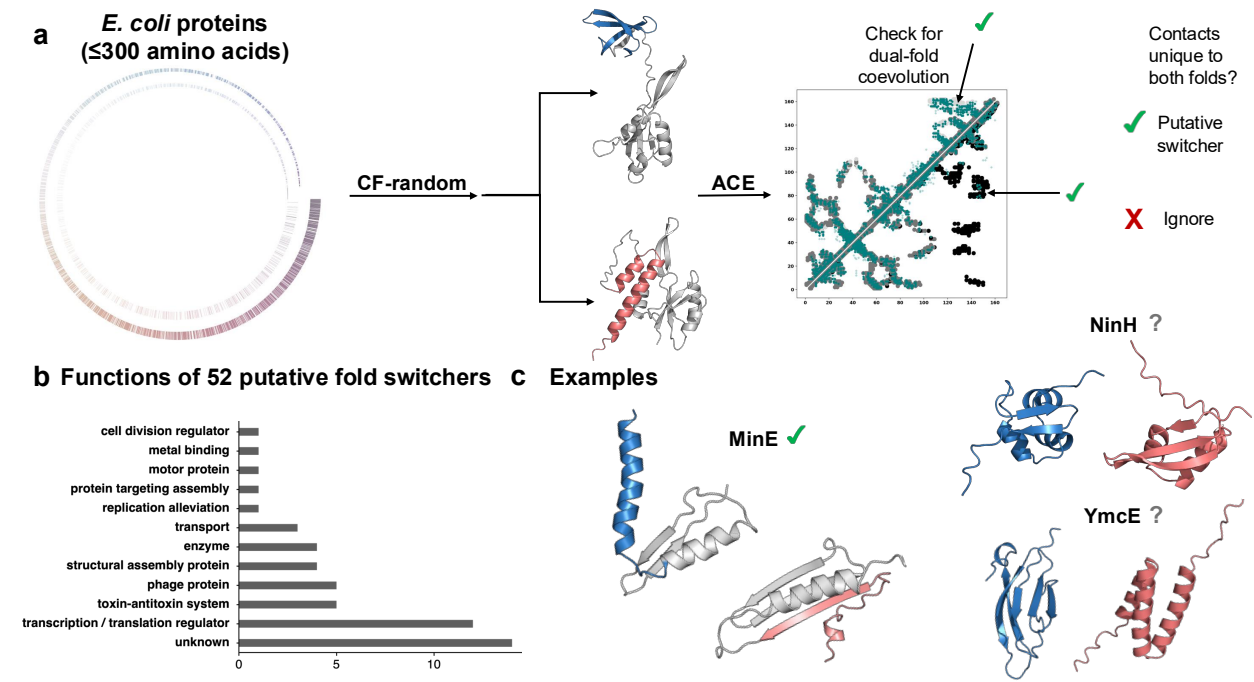

**Figure 4. Predicting *E. coli* proteins that may switch folds.** (A). 2126 *E. coli* and phage proteins were run through CF-random to test whether they switch folds (59). Seashell-like image represents these proteins by length; the inner circle represents 1111 candidates for which sufficiently deep MSAs could not be generated, and outer, the 2126 proteins that were then run through CF-random. If two or more distinct conformations were identified, such as in the case of the successfully identified fold-switching *E. coli* protein, RfaH, we tested for dual-fold coevolution using ACE (95). If coevolutionary evidence for both folds was found, the protein was considered a putative fold switcher. Light gray/black contacts on upper/lower diagonals are unique to CF-random predicted dominant/alternative conformations. Teal contacts are from ACE. Medium gray contacts are common to both folds. (B). Putative fold-switching proteins are involved in diverse functions. (C). Examples of putative hits. CF-random correctly identified the fold-switching protein MinE from its thousands of candidates, indicated by green check. NinH is transcription factor protein that may undergo an α-helix-to-β-sheet transition, and YmcE is a bacterial antitoxin predicted to assume two different folds with lower confidence; both YmcE and NinH are putative switchers indicated by gray question marks.

Nevertheless, CF-random was developed to blindly search whole proteomes for new fold switchers. When applied to thousands of *Escherichia coli* proteins, it predicted that up to 5% may switch folds (**Figure 4**). To our knowledge, CF-random is among the first tools capable of reliably generating plausible 3D alternative conformations across large datasets and winnowing the number of possible alternative conformations to a few, offering a scalable strategy for de novo discovery of fold-switchers. Supporting the widespread role of fold-switching proteins in nature, a support vector machine-based approach with a Matthews Correlation Coefficient of 0.7 recently suggested that 1-5% of globular (non-disordered) proteins from dozens of organisms may switch folds (102).

*5.2. Designing new fold switchers*

Near the end of his life, Richard Feynman wrote, "What I cannot create, I do not understand" (112). The unique ability of fold-switching proteins to toggle between functional states in response to specific stimuli, such as temperature or pH, has inspired approaches to understand fold switchers by designing them. Nearly 20 years ago, Ambroggio and Kuhlman reported the first computationally designed fold-switching protein, Sw2, that switched between zinc-finger and trimeric coiled-coil folds in response to pH or the presence of transition metals (5, 6). Three years later, Bryan and Orban combined directed evolution with rational design to develop a set of proteins with 98% sequence identity but different folds (3) and recently built on this success to engineer the temperature-sensitive fold switcher Sa1 V90T (101). Besides these, few approaches (115) have been developed to engineer fold-switching proteins until recently.

The success of deep learning models has provided new impetus to addressing the difficult problem of multi-state protein design. Last year, the Baker lab used a diffusion-based model to design an α/β protein whose individual halves each assumed α-helical structures (65). This impressive result highlights remaining challenges in characterizing fold switchers: though the CD spectra of the individual protein halves were consistent with all helical structures, their structures were not determined experimentally, and some of their NMR spectra showed broad lines, possibly indicating unstable alternative conformations. Deep-learning methods have also recently been used to engineer proteins with hinge-like motions (84) and dynamic proteins (44). These advances suggest that the future for computational designs of new fold-switching proteins is bright.

## 6. Future opportunities

Fold-switching proteins challenge conventional wisdom and raise fundamental unanswered questions about protein structure, biophysics, and evolution. Recent work has begun to address these questions by characterizing the stabilities and interconversions of fold switchers, finding coevolutionary signals unique to both folds, identifying mutational pathways that cause proteins to switch between one fold and another, and developing predictive methods to identify new fold switchers from genomes and define their biological scope. Nevertheless, barriers to progress remain.

Discovery of new fold switchers–rather than homologs of known ones–through reliable computational methods would be a major advance to the field. To our knowledge, this has not been achieved yet. It requires computational methods that reliably predict new fold switchers from their amino acid sequences sequence. Though current methods identify known fold switchers, they miss others, and their ability to predict new fold switchers remains unproven. These predictions will ultimately need to be tested experimentally. Further, more information about what triggers a predicted switch may be needed for experiments to be successful.

Experimental methods that screen for fold switchers among a high volume of predicted targets are needed. Ideally, these methods would not require laborious protein purification. Currently, there is no one-size-fits-all solution. In-cell Förster Resonance Energy Transfer has worked in some cases (80), but it will not work for fold-switching proteins whose distinct conformations have similar end-to-end distances. Hydrogen-deuterium exchange mass spectrometry has successfully identified other sorts of conformational changes (23), but it is slow (~7 hours/sample) and may require pure protein. NMR has fostered discoveries of several fold switchers (80), but it requires large amounts of pure labeled protein. For the foreseeable future, discovery of new fold switchers may require combining or trying several methods.

Finally, fundamental questions about the sequence-structure relationship remain unanswered. What sequence features cause one protein to switch folds when its homolog doesn't? How do the sequence changes caused by alternative splicing and SNPs cause proteins to dramatically switch their folds? Methods designed to characterize protein function and energy landscapes on a large scale (36, 71) may help answer these questions. By doing so, we may better understand not only how fold switching works but also how mutations affect energy landscapes in general. We hope that future discoveries and characterizations of fold-switching proteins help to answer these questions.

**TERMS AND DEFINITIONS**

**Metamorphic proteins** reversibly interconvert between folds with different secondary and/or tertiary structures, enabling functional versatility and response to environmental stimuli.

**Fold-switching proteins** remodel their secondary and/or tertiary structures in response to external stimuli such as changes in temperature, pH, binding to a partner, or oligomerization. Fold switching can be reversible or irreversible.

**Intrinsically disordered proteins (IDPs)** do not spontaneously fold into stable, well-defined 3D structures. Instead, they remain dynamically flexible, rapidly shifting among a range of conformations that span from extended coils to compact globules.

**The energy landscape** of a protein statistically describes the potential energy surface of its structure. For most globular proteins, this landscape is funnel-shaped, with a single deep energy well guiding the polypeptide chain toward a unique, thermodynamically stable conformation or native fold. However, IDPs feature a flat energy landscape, while fold-switching proteins have multiple metastable energy wells, implying more than one native state in their landscape.

The **Monte Carlo** (**MC**) method is a set of computational techniques that utilize random numbers to derive approximate solutions for mathematical problems; among other applications, it is used to sample a protein's conformational space.

**Replica exchange MD (REMD)** - Replica-exchange simulated tempering, also known as parallel tempering, is a MC simulation technique used to efficiently sample the configuration space of complex systems, such as proteins. Multiple replicas of the system are simulated in parallel at different temperatures, enabling exploration across a range of energy landscapes.

**Replica exchange with tunneling (RET)** is an advanced variant of **Replica exchange** designed to accelerate sampling, particularly for systems with rugged energy landscapes like protein misfolding or aggregation. This method incorporates transitions not only between temperatures but also between different levels of system resolution, typically switching between coarse-grained and fine-grained representations.

**Markov models -** Markov chain MC simulation (MCMC) is a method that generates states from a general probability distribution.

**Structure-based models (SBMs)** simplify the potential energy function by focusing on interactions that are present in the native, folded structure of the protein. In single-basin SBMs, the energy landscape is shaped as a funnel biased toward a single native conformation, typically derived from the contact map of the folded structure, derived from either experimentation or modeling.

**Alternative splicing** is a process during gene expression in which different protein-coding regions of RNA are joined together in different ways from a pre-mRNA transcript. This generates multiple distinct mRNA and, consequently, protein variants from one gene, increasing the diversity of proteins in a cell.

**Post-translational modifications (PTMs)** are chemical changes made to a protein after it has been synthesized (translated by ribosomal assembly). Common modifications include phosphorylation, glycosylation, or acetylation, and can alter a protein's structure, stability, localization, or activity, and are crucial for regulating protein function.

**Single-nucleotide polymorphisms (SNPs)** are the most common type of genetic variation, involving a change of a single nucleotide in the DNA sequence. SNPs can affect protein structure and function, influence disease susceptibility, or be used as genetic markers.

**Evolved fold switching** refers to significant structural changes in a protein or organism that occur over evolutionary time in response to stepwise mutation.

**Ancestral reconstruction (ASR)** is a technique used to infer the probable characteristics or sequences of ancestral organisms represented by the internal nodes of a phylogenetic tree. By analyzing the traits of living species and considering their evolutionary relationships, ASR uses models of trait change to estimate what the common ancestors of these species were likely to be.

**A protein family** is a group of proteins that share a common evolutionary origin, reflected by similar sequences or structures. Members of a protein family typically have significant sequence similarity and often perform related biological roles.

**A protein subfamily** is a smaller group (clade) within a protein family, whose members are even more closely related to each other by sequence, structure, or function. Subfamilies often arise due to gene duplications and subsequent diversification, and they typically share more specialized features or functions compared to the broader family.

**A protein superfamily** is a large grouping of distantly related protein families that share structural or functional features, even if their sequences have diverged significantly. Proteins in a superfamily usually have a common structural framework or evolutionary origin but may perform a wide range of biological functions.

**Coevolutionary signals** arise from correlated mutations among related protein sequences and almost always correspond to with pairs of amino acids in direct contact; also known as evolutionary couplings.

**Large language models (LLMs)** are advanced artificial intelligence (AI) systems designed to understand and generate human-like text by learning patterns and rules from enormous datasets containing billions of words. These models perform a wide range of natural language processing (NLP) tasks, such as answering questions, summarizing text, translating languages, and generating written content. In biophysics and related fields, LLMs are increasingly being used to analyze scientific literature, assist with protein sequence analysis, predict protein secondary structure or folds from sequence data, and generate hypotheses by identifying patterns in biological texts and datasets.

**SUMMARY POINTS**

1. Fold-switching proteins challenge conventional wisdom about protein structure, biophysics, and evolution and are gaining scientific traction, underscoring the focus on understanding the complexities of protein folding and structure.

2. The limited presence of fold-switching proteins in structural databases such as the Protein Data Bank highlights limitations of contemporary experimental techniques in solving structures of multi-conformational proteins. The biological scope of fold-switching proteins remains uncertain.

3. Fold-switching proteins have multi-well energy landscapes that are impacted by environmental factors such as temperature. Fold-switching transitions are slow (on the order of seconds or slower) and can occur through intermediates.

4. Evolution has selected for both folds of many fold-switching proteins, and new folds can emerge in protein families mediated by stepwise mutation.

5. A combination of experimental and computational methods has been used to understand the fold-switching phenomenon, pushing the field forward to develop or combine clever techniques to characterize fold-switchers.

6. New fold-switching systems have been successfully designed that can stably adopt multiple, distinct, well-defined structures and reversibly interconvert between them.

**FUTURE ISSUES**

1. Correctly predicting novel fold switchers from amino acid sequences remains a major challenge, as current computational tools can sometimes identify known examples but haven't identified new ones yet. Once this barrier is crossed, the biological scope of fold switching will become clearer.

2. Experimentally testing of predictions is essential but hindered by limitations of current methods, which are often slow, labor-intensive, or not broadly applicable. Techniques like in-cell FRET, HDX-MS, and NMR each have benefits and drawbacks, suggesting that a combination of approaches may be necessary.

3. Testing predictions requires additional biological knowledge. A prediction could be correct, but if its trigger is not known, fold switching will be difficult to observe.

4. Fundamental questions persist, including what sequence features enable fold switching and how genetic variations influence folding. Large-scale studies of protein functions and energy landscapes may offer insights.

**Acknowledgements**

L.L.P. thanks Joshua Porter for helpful discussions. This work utilized resources from the NIH HPC Biowulf cluster (https://hpc.nih.gov/) and was supported by the Intramural Research Program of the National Library of Medicine, National Institutes of Health (LM202011, L.L.P.)

**Citation information**

When citing this paper, please use the following: Chakravarty D, Porter LL. 2026. Fold-switching Proteins. Annu. Rev. Biophysics. 55: Submitted. DOI: 10.1146/annurev-biophys-022924-012038